\documentstyle[11pt,newpasp,twoside]{article}
\def\edcomment#1{\iffalse\marginpar{\raggedright\sl#1\/}\else\relax\fi}
\reversemarginpar

\begin{document}
\phantom{x} \vskip -0.3in \phantom{x}

\title{            Counts of Low-redshift SDSS Quasar Candidates      }

\phantom{x} \vskip -0.5in \phantom{x}
\author{\v{Z}eljko Ivezi\'{c}$^{1,2}$, Robert H. Lupton$^1$, David E. Johnston$^3$, 
Gordon T. Richards$^1$, Pat B. Hall$^{1,4}$, David Schlegel$^1$, Xiaohui Fan$^5$, 
Jeffrey A. Munn$^6$, Brian Yanny$^7$, Michael A. Strauss$^1$, Gillian R. Knapp$^1$, 
James E. Gunn$^1$, Donald P. Schneider$^8$}

\affil{$^1$ Princeton University, Princeton, NJ 08544}
\affil{$^2$ H.N. Russell Fellow, on leave from the University of Washington}
\affil{$^3$ University of Chicago, 5640 S. Ellis Ave., Chicago, IL 60637}
\affil{$^4$ Pontificia Univ. Cat\'{o}lica de Chile, Casilla 306, Santiago 22, Chile}
\affil{$^5$ Steward Observatory, The University of Arizona, Tucson, AZ 85721}
\affil{$^6$ U.S. Naval Observatory, P.O. Box 1149, Flagstaff, AZ 86002}
\affil{$^7$ FNAL, P.O. Box 500, Batavia, IL 60510}
\affil{$^8$ The Pennsylvania State University, University Park, PA 16802}

\begin{abstract}
We analyze the counts of low-redshift quasar candidates selected using
nine-epoch SDSS imaging data. The co-added catalogs are more than 1 mag 
deeper than single-epoch SDSS data, and allow the selection of 
low-redshift quasar candidates using UV-excess and also variability techniques.
The counts of selected candidates are robustly determined down to $g$=21.5.
This is about 2 magnitudes deeper than the position of a change in the slope of the 
counts reported by Boyle et al. (1990, 2000) for a sample selected by UV-excess, 
and questioned by Hawkins \& Veron (1995), who utilized a variability-selected sample. 
Using SDSS data, we confirm a change in the slope of the counts for both UV-excess 
and variability selected samples, providing strong support for the Boyle 
et al. results.
\end{abstract}

\vskip -0.55in
\phantom{x}
\section{A Controversy About Quasar Counts}
\vskip -0.10in
\phantom{x}

The quasar luminosity function, $\Phi(L)$, provides fundamental information
about their nature. Boyle et al. (1990, 2000) found that $\Phi(L)$ for quasars
at redshifts $z<2.3$ resembles a ``broken'' power law which becomes {\it flatter} 
at the faint end. They also demonstrated that the ``break'' luminosity at which 
the slope changes increases with redshift, and between $z=0.8$ and $z=2.2$ 
becomes more luminous by 1.9 mag. A peculiar aspect of this result is that the apparent 
magnitude corresponding to the ``break'' luminosity changes by only 0.8 mag. 
over this redshift range ($B$=18.6--19.4; that is, the differential distribution of 
quasar apparent magnitudes has a shape very similar to the shape of their luminosity 
function). Further, this apparent magnitude 
range is only $\sim$ 1-2 mag. above the photographic plate limit used for selection. 
Therefore, it is possible that the flattening of the luminosity function is due 
to a systematic underestimate of the increasing incompleteness at the faint end of 
the UV-excess selection technique employed by Boyle et al. Indeed, Hawkins \& Veron (1995), 
using a variability-selected sample with 300 objects, argued that ``The luminosity 
functions for redshifts of less than 2.2 show a featureless power law, with no sign 
of  a `break'.''

\vskip -0.3in
\phantom{x}

\section{ Can SDSS Resolve the Controversy? }

The $z<3$ SDSS (York et al. 2000) spectroscopic quasar sample is limited to $i<19.1$, and 
thus not deep enough to reliably constrain the faint end of the luminosity function.
The color selection by UV excess from the SDSS photometric sample 
(essentially based on SDSS $u-g$ color, see Richards et al. 2002 and
references therein) is limited by the photometric accuracy in the $u$ band 
to $u \sim 21$, which would be just barely sufficient to detect a change of 
the slope of counts at $i \sim 20$. However, the coadding of SDSS multi-epoch 
photometry produces $>$1 mag. deeper $u$-band measurements at a given accuracy, 
and allows a robust identification of UV-excess selected candidates to $u \sim 22$. 
Furthermore, multi-epoch data can also be used to select quasar candidates by 
variability, using techniques similar to those employed by Hawkins \& Veron. 
This is the first time that it is possible to contrast the conflicting results 
of Boyle et al. and Hawkins \& Veron using the same homogeneous data set.  

\begin{figure}[t]
\plotfiddle{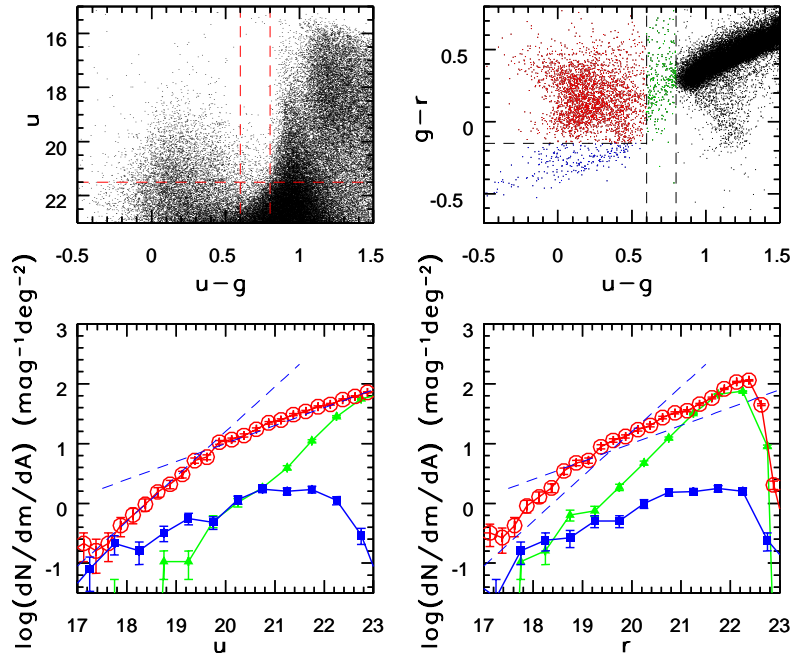}{8cm}{0}{55}{55}{-170}{-110}
\caption{The top panels show the distribution of $\sim$220,000 point sources 
from 60 deg$^2$ of sky with 9 SDSS imaging observations in representative
SDSS color-magnitude and color-color diagrams. The bottom two panels show 
differential counts for sources from three regions in the $g-r$ vs. $u-g$ 
color-color diagram: quasar candidates (circles, $u-g < 0.6$, $-0.15 <g-r< 0.8$), 
hot stars (triangles, $u-g < 0.6$, $<g-r<-0.15$), and quasar-star transition 
region (squares, $0.6 <u-g< 0.8$). The dashed lines, with slopes of 0.75 and 0.30, 
are added to guide the eye (same lines in both panels).
}
\vskip -0.2in
\end{figure}

%\vskip -0.55in
%\phantom{x}
%\vskip -0.55in

\section {Color Selection}
\vskip -0.25in \phantom{x}

We selected low-redshift quasar candidates by requiring $u-g < 0.6$ and
$-0.15 <g-r< 0.7$ (based on the criteria of Richards et al. 2002). Their counts,
and counts for two control samples, are shown in Fig. 1. The differential
counts of selected candidates rise steeply (log($N$) vs. $m$ slope $\sim$0.75) 
at the bright end ($u<19.5$, or $r<19$), and then the slope changes to $\sim$0.30. 
This behavior is in agreement with the results of Boyle et al., who used a similar
selection technique. It is worth emphasizing that SDSS photometric catalogs 
based on digital data are much more accurate (0.02 mag. errors, see Ivezi\'{c} 
et al. 2003) than the catalogs used by Boyle et al. which were derived from a photographic 
survey. In particular, the gap between the distribution of quasars and more
numerous blue thick disk and halo stars 
(centered on $u-g\sim0.7$, see the top left panel in Fig. 1) cleanly separates 
the two populations.

We note that some of the UV-excess selected candidates are not quasars: the results
of SDSS spectroscopic follow-up ($i<19$) indicate that $\sim20\%$ are stars
(dominated by white dwarfs).
This correction, however, has no effect on the conclusion that the slope of
differential counts changes because there are about 10 times fewer quasar 
candidates at $r\sim21.5$ than predicted by extrapolating the counts for 
$r<19$. Similarly, star/galaxy separation is not an issue: even if resolved 
sources were added to the point sources, the counts would be qualitatively 
unchanged (i.e. the UV-excess sources are heavily dominated by point 
sources).

\vskip -0.35in \phantom{x}
\section{ Variability Selection }
\vskip -0.25in \phantom{x}

Quasars are variable sources with amplitudes of several tenths of a magnitude
for time scales longer than a few months (e.g. Hawkins \& Veron 1995). The top 
left panel in Fig. 2 shows the correlation between 
variability in the $g$ band, $|\Delta g|$, and $u-g$ color for point sources 
observed twice two years apart. A large majority of variable sources, defined 
by $|\Delta g|>0.15$, show UV excess ($84\%$ for $u<21.5$). Even without 
accounting for the fact that some variable sources are stars (which would further 
increase this fraction), this high fraction implies that UV-excess and variability selection 
techniques yield similar samples (see the two bottom panels in Fig. 2). 
Alternatively, 47\% of sources selected by UV-excess have $|\Delta g|>0.15$. 
It is therefore not surprising that the counts of variability-selected quasar 
candidates, which we define as {\it all} sources with
$\Delta(g)>0.15$, also show a change of slope at $r\sim 19$.  
Of course, this selection could be further refined; for example, RR Lyrae
stars, which dominate the variable star population, can be easily removed
due to their distinctive colors and much shorter variability time scale 
(Ivezi\'{c} et al. 2000). Such improvements would only further increase 
the similarity of differential counts for UV-excess and variability selected 
quasar candidates.

\begin{figure}[th]
\plotfiddle{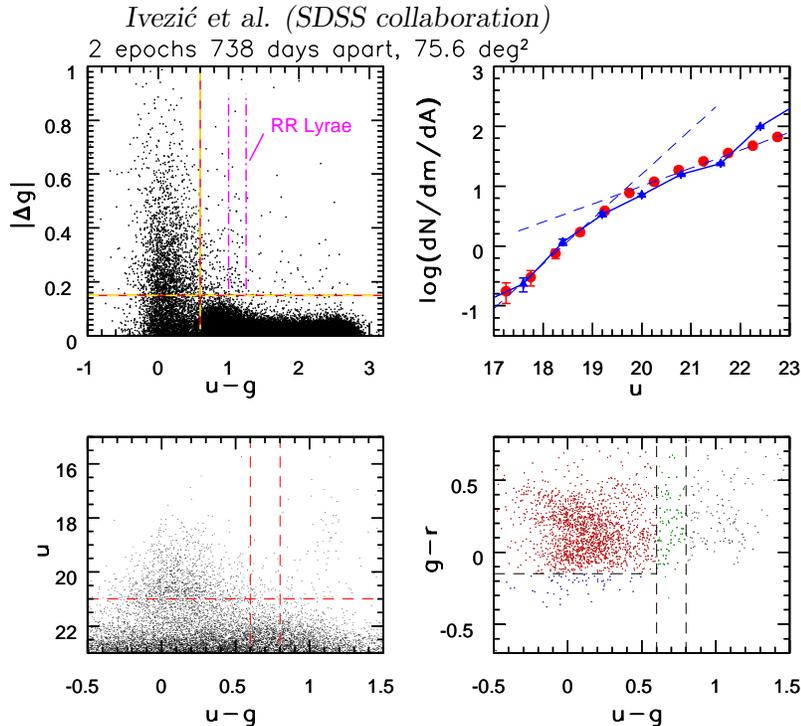}{7.8cm}{0}{55}{55}{-170}{-110}
\caption{The top left panel shows the difference in $g$ magnitude as a function of $u-g$ 
color for 47,116 points sources with mean $u<21.5$ observed twice $\sim2$ years 
apart. The vertical full line illustrates the UV-excess selection of low-redshift 
quasar candidates from Figure 1 ($u-g<0.6$), and the horizontal line shows the 
adopted limit for the selection of variable sources ($|\Delta g|>0.15$ mag). 84\% of 
variable sources show UV-excess, and 47\% of UV-excess selected sources have 
$|\Delta g|>0.15$ mag. The top right panel compares the counts of sources 
selected by UV-excess (circles, same as bottom left panel in Fig. 1), and
those selected by variability (triangles, renormalized at the bright end, $u=19$).
The dashed lines are the same as those in bottom left panel in Fig. 1.
The bottom two panels show the distribution of variability-selected sources in 
color-magnitude and color-color (only $u<21.5$) diagrams (compare to the top
panels in Fig. 1). 
}
\vskip -0.25in
\end{figure}

\vskip -0.35in
\phantom{x}
\section { Discussion  }
\vskip -0.25in \phantom{x}

The counts of quasar candidates selected by both UV-excess and variability 
techniques using multi-epoch SDSS imaging data change slope around $r=19$. 
This result provides strong support for the quasar luminosity function and 
its evolution advocated by Boyle et al., and demonstrates that both techniques
produce similar samples when applied to high-quality data.

\vskip -0.15in
\phantom{x}

{\small 
Funding for the creation and distribution of the SDSS Archive has been provided by 
the Alfred P. Sloan Foundation, the Participating Institutions, the National Aeronautics 
and Space Administration, the National Science Foundation, the U.S. Department of Energy, 
the Japanese Monbukagakusho, and the Max Planck Society. The SDSS Web site is 
http://www.sdss.org/. 
}
\vskip -0.4in
\phantom{x}

\end{document}